\documentclass[twocolumn,prd,superscriptaddress,showpacs,amsmath,amssymb]{revtex4-1}

\usepackage{graphicx}
\usepackage{dcolumn}
\usepackage{bm}
\usepackage{multirow}

\begin{document}

\title{Quenching and Channeling of Nuclear Recoils in NaI[Tl]: \\ Implications for Dark Matter Searches}

\def\UC{Kavli Institute for Cosmological Physics and Enrico Fermi Institute, University of Chicago, Chicago, IL 60637, USA}

\author{J.I.~Collar} \email{Electronic address: collar@uchicago.edu}\affiliation{\UC}
\date{\today}

\begin{abstract}
A new experimental evaluation of the quenching factor for nuclear recoils in NaI[Tl] is described. Systematics affecting previous measurements are addressed by careful characterization of the emission spectrum of the neutron source, use of a small scintillator coupled to an ultra-bialkali high quantum efficiency photomultiplier, and evaluation of non-linearities in the electron recoil response via Compton scattering. A trend towards a rapidly diminishing quenching factor with decreasing sodium recoil energy is revealed. Additionally, no evidence for crystal lattice channeling of low-energy recoiling ions is found in a scintillator of known crystallographic orientation.  A discussion on how these findings affect dark matter searches employing NaI[Tl] (e.g., DAMA/LIBRA) is offered.
\end{abstract}



\maketitle

\section{Introduction}

The interpretation of dark matter search data depends critically on a complete  understanding of detector response to nuclear recoils. These are expected to be induced via elastic scattering of dark matter particles off target nuclei \cite{goodman}. The modest recoil energies anticipated from Weakly Interacting Massive Particle (WIMP) interactions complicates this necessary labor of detector characterization. In recent times, a number of potential signals have been reported by several such searches. These anomalies \cite{DAMA,cogent,cresst}, together with claims of exclusion of the relevant WIMP phase space by other experiments \cite{xenon,cdms}, have underlined the importance of proper detector characterization prior to the reporting of positive or negative results. 

Nuclear recoils (NR) induce a diminished response in target materials, when compared to electron recoils (ER) of the same energy. In scintillators, this is mainly due to the small fraction of the recoil energy ($E_{r}$) that is transferred to electron excitation: an energy-dependent quenching factor ($Q$) can be defined as the ratio of the light yield from a nuclear recoil to that from an electron recoil of the same energy. Nuclear recoil energies (typically expressed in units of keVnr) therefore correlate to a smaller observable electron equivalent energy (keVee) through this quenching factor. The energy scale is typically calibrated using monochromatic gamma rays, leading to a satisfactory knowledge of the electron recoil response. However, in order to predict the response to WIMP-induced nuclear recoils, a dedicated characterization of the quenching factor is necessary. This is typically accomplished using neutron-induced nuclear recoils.

Quenching factor measurements can be divided into two groups. The first involves use of sources generating a broad spectrum of neutron energies, such as AmBe or $^{252}$Cf. A comparison between the simulated and observed response to the source is used to generate a best-fitting quenching factor, often approximated as being energy-independent \cite{damaquenching, fushimi}. If an attempt is made to include the dependence on energy, uncertainties in the simulation and systematic effects in the measurement are forced into the derived $Q(E_{r})$. This is specially grave at the smallest energies investigated (few keVnr) \cite{xecritique1,xecritique2}. A second more reliable method, yet not entirely devoid of its own issues \cite{xecritique1}, involves monochromatic neutron sources. The detection of the scattered neutron at a fixed angle from the incoming neutron beam leads to a precise knowledge of the recoil energy deposited in the target material under test \cite{chag,jagemann,simon,gerbier,tovey,spooner}. 

In the particular case of NaI[Tl] scintillators like those used by the DAMA/LIBRA collaboration \cite{DAMA}, the first method has been reported to yield a sodium recoil quenching factor $Q_{\rm Na} = 0.30 \pm 0.01$
averaged over 6.5 to 97 keVnr \cite{damaquenching}. For iodine recoils this was $Q_{\rm I} = 0.09 \pm 0.01$
over the range 22 to 330 keVnr \cite{damaquenching}. In \cite{fushimi} the  values found were 
$Q_{\rm Na} = 0.4 \pm 0.2$ and $Q_{\rm I} = 0.05 \pm 0.02$, averaged over 5-100
keVnr and 40-300 keVnr, respectively. The second method yields values in principle compatible with these. However, in the most recent measurement of this second type \cite{chag}, a trend towards an increasing  $Q_{\rm Na}$ with decreasing recoil energy is observed down to the lowest measured energy, $E_{r}=$10 keVnr. This increase is also predicted by a semi-empirical treatment of the scintillation mechanisms at play \cite{tretyak}. Indeed, the kinematic threshold \cite{kinematic} below which this $Q_{\rm Na}$ would be expected to decrease should appear at a considerably smaller $\sim$2.5 keVnr for sodium recoils in NaI[Tl] \cite{danjuan}.  In contrast to this possibility, other scintillators such as liquid xenon (LXe) are expected to display a decreasing quenching factor starting already below kinematic thresholds at few tens of keVnr \cite{danjuan}. 

The channeling of recoiling ions within a crystal lattice constitutes an additional source of uncertainty specific to single crystal scintillators such as NaI[Tl]. If this process is present at the few keVnr energies of interest, it can result into a considerable displacement of the region of WIMP parameter space (mass, coupling) able to explain DAMA/LIBRA observations, away from values already severely constrained by other experiments \cite{prochannel}. A recent theoretical reanalysis of this possibility \cite{conchannel} concluded that while ion channeling is a process well-established at higher energies, it should play a negligible role for dark matter searches.

The new measurement of $Q_{\rm Na}$ and $Q_{\rm I}$ in NaI[Tl] presented here attempts to address the uncertainties described above with the introduction of several improvements, delineated in the next section. These include the use of a scintillator with a (partly) known crystallographic orientation. This allowed a first experimental test of the possibility of channeling at few keVnr. An increasing $Q_{\rm Na}$ towards low $E_{r}$ and/or non-negligible ion channeling would result in a common region of interest in WIMP phase space for the DAMA, CoGeNT and CRESST anomalies \cite{hooperetal,kelso}. As discussed below, neither of these effects was observed. Based on present results, the energy-independent values  $Q_{\rm Na} = 0.3$, $Q_{\rm I} = 0.1$ that are typically adopted in the interpretation of the DAMA/LIBRA annual modulation should be held suspect. The last section in this paper briefly discusses the impact of these findings on the interpretation of DAMA/LIBRA results, and their compatibility with other searches. 

\section{Present Improvements to the Method}

The experimental setup delineated in this and the following sections was designed to address several sources of systematic error affecting previous measurements. These measures can be listed:

\begin{figure}[h]
\includegraphics[width=0.46\textwidth]{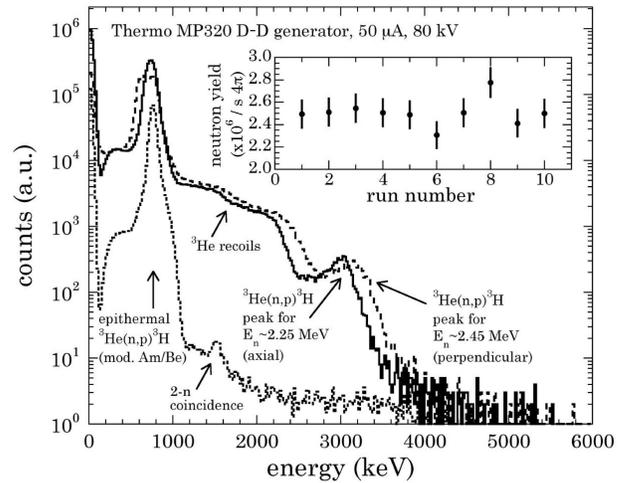}
\caption{\label{fig:CoGeNT-Traces} Measurement of $E_{n}$ using a C-S spectrometer (see text). Pulse-shape analysis \protect\cite{he3} was unnecessary, in view of the clear separation between the $^{3}$He(n,p)$^{3}$H peak and $^{3}$He recoils. The conventional response to epithermal neutrons from a moderated Am/Be source is shown for reference. The NaI[Tl] scintillator was positioned in front of the MP320 end nearest to the target plane, therefore sustaining a $E_{n}=$2.25 MeV irradiation. {\it Inset}: measurements of neutron yield using the rate under the fast-neutron peak in a 1 c.c.\ $^{6}$LiI[Eu] scintillator \protect\cite{li6}. A simulation of the $^{6}$Li(n,$\alpha$)$^{3}$He reaction rate using MCNP converts this measured rate into the isotropic point source-equivalent yield depicted, in good agreement with D-D generator specifications. Multiple measurements were made to monitor the absence of deuterated target depletion during neutron scattering runs. Error bars are statistical only, with the dominating systematic being the reproducibility in the positioning of the $^{6}$LiI[Eu]  crystal.}
\end{figure}

\begin{list}{$\circ$}{}  

\item New photomultipliers with enhanced quantum efficiency (Hamamatsu's Ultra-Bialkali, UBA) became commercially available since the last previous quenching factor measurement in NaI[Tl] \cite{chag}. Their use here results into a light yield of up to 20 photoelectrons (PE) per keVee, a factor of $\sim$4 larger than in \cite{chag}, i.e., an energy threshold lower by the same factor. The reader is referred to the discussion around Fig.\ 4 in \cite{xecritique1} for a detailed explanation on how an insufficient light yield can result into artificially large values of $Q$ at the lowest recoil energies measured. Taking as a reference the $E_{r}\sim$10 keVnr at which the present measurement starts to suffer from such threshold effects, and scaling by the presently achieved increase in light yield, it is possible to conclude that all previous measurements of $Q_{Na}$ below $E_{r}\sim$40 keVnr have been disturbed to some extent by the artifact described in \cite{xecritique1}. In contrast to this, only the lowest $E_{r}=$8 keVnr data point in this work is expected to be somewhat affected. 

\item The small dimensions of the NaI[Tl] crystal presently used (17$\times$17$\times27\!$ mm) were sanctioned via MCNP-PoliMi Monte Carlo simulation \cite{polimi} to ensure that multiple-scattering played a negligible role, comprising just $\sim$10.5\% of total neutron interactions. This is to be compared with the considerably larger crystals used in previous measurements (e.g., 5 cm diameter by 5.4 cm long in \cite{chag}). Multiple scattering in oversized detectors creates a background continuum able to encumber single-scattering signals. This can result in systematic effects leading to artificially large values of $Q$. A recent example of how a reduction in detector size can help alleviate this background and eliminate such systematics can be found in a comparison of the results from \cite{manamana} with those in \cite{plante}.

\item The response to low-energy electron recoils in NaI[Tl] is known to exhibit a measurable non-linearity \cite{nonprop1,nonprop2}, reaching a maximum in light yield at around 10 keVee. While this has been acknowledged in a few previous quenching factor measurements \cite{chag,gerbier}, this response is typically simply normalized to that at $\sim$100 keVee, and approximated as linear. This action, on its own, results into overestimating $Q_{Na}$ in the region of energy next to DAMA/LIBRA's threshold ($\sim$2 keVee) by approximately 15\% (the precise electron recoil energy scale in DAMA/LIBRA can be stablished through a convenient line at 3.2 keVee due to $^{40}$K contamination). In this work the response to electron recoils is measured throughout the range 2-50 keVee via Compton scattering. Emphasis is placed on minimizing alterations of the experimental setup and analysis protocol between electron scattering and neutron scattering measurements. The quenching factor measured here folds in the effect of any non-linearities by taking the ratio between light yields for NR and ER of the same energy, at all energies studied. Unfortunately, this optimal approach to extracting $Q $ is not widely used: the extent to which a large non-linear response to low-energy electron recoils in LXe \cite{lxee} might affect the interpretation of dark matter experiments using that target is an open question.

\item Most monochromatic neutron sources exhibit some energy dependence on the emission angle, measured from the direction of the projectile incident on the target sustaining the reaction. In the case of a D-D "neutron gun" (deuterium onto a deuterated target) like the Thermo MP320 generator employed here, this is known to be $E_{n}\sim$2.8 MeV in the forward direction, $E_{n}\sim$2.4 MeV at 90$^{\circ}$, and $E_{n}\sim$2.2 MeV at 180$^{\circ}$, for an accelerating potential of 80 kV \cite{nbook}. Information about the sense of the deuterium projectile along the longitudinal axis of the gun was however not available from the manufacturer. The neutron energy emitted towards the NaI[Tl] scintillator ("axial" in Fig.\ 1) was measured using a large-volume $^{3}$He counter (9.2 cm diameter, 31 cm long). A counter this large is able to contain the full projected range of protons generated by the $^{3}$He(n,p)$^{3}$H reaction. This allows for low-efficiency, high-resolution spectroscopy of monochromatic fast neutron sources, for which a full-energy deposition peak is detected at $E_{n}+764$ keV, where 764 keV is the Q-value of this exothermic reaction \cite{he3} (Fig.\ 1).  $^{3}$He counters used for this unconventional application are referred to as Cuttler-Shalev (C-S) spectrometers \cite{he3}. This characterization allowed to remove an uncertainty in $E_{n}$ that may have affected previous use of the Thermo MP320 \cite{dan1,dan2}. Additional measurements of the stability of neutron yield were performed using a small $^{6}$LiI[Eu] scintillator (Fig.\ 1).
 
\end{list}

\begin{widetext}

\begin{figure}[h]
\includegraphics[width=1.\textwidth]{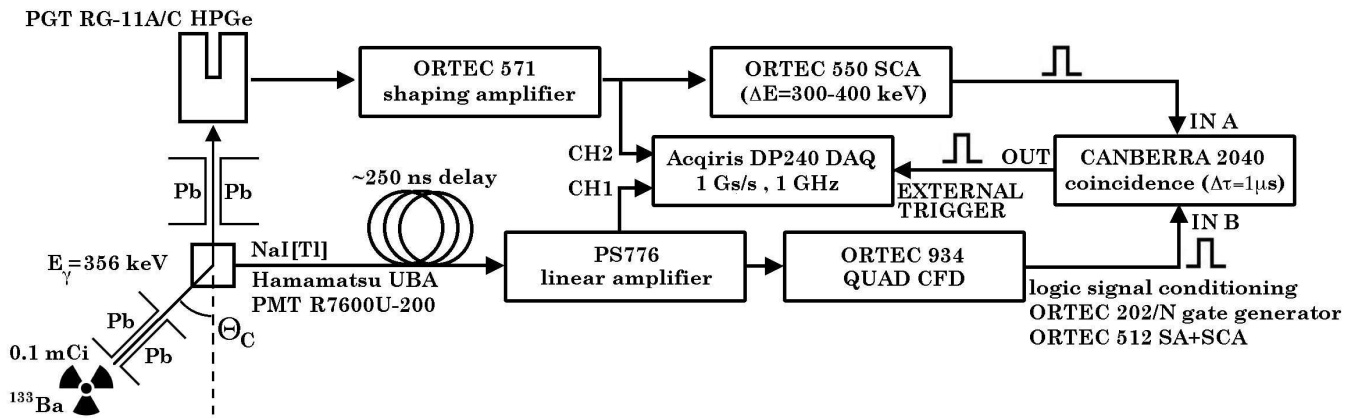}
\caption{\label{fig:electronics} Experimental setup used for Compton scattering measurements (see text). }
\end{figure}

\end{widetext}

\section{Compton Scattering Measurement}

Fig.\ 2 displays the setup used to measure the response of NaI[Tl] to low-energy electron recoils. A pencil beam (0.6$^{\circ}$ aperture) of collimated  356 keV gammas from a 0.1 mCi $^{133}$Ba source bathed the NaI[Tl] scintillator. Gamma source and lead collimator were mounted onto a goniometric stage used to select the Compton scattering angle, $\Theta_{C}$. The square cross-section (17$\times$17$\times27\!$ mm) Czochralski-grown NaI[Tl] single crystal was obtained from Proteus/Amcrys. It contained approximately 700 ppm of thallium dopant. This is similar to DAMA/LIBRA's quoted $\sim$0.1\% \cite{dopant} (small variations in dopant concentration are not expected to affect quenching factor measurements \cite{kims}). A request was made to the manufacturer to have two of its side faces be cut perpendicular to the [100] crystal growth axis (Fig.\ 5). The scintillator was vertically mounted onto a Hamamatsu ultra-bialkali (UBA) R7600U-200 photomultiplier (PMT), using optical RTV as a couplant. PMT and base were mounted onto a second goniometric stage \cite{stage} used to rotate them around their vertical during ion channeling measurements (Fig.\ 5, section VI). Both were surrounded by a sheath of mu-metal to avoid magnetic field disturbances to the PMT gain upon rotation. The precision of the angular alignment of all elements involved is estimated at $\sim1^{\circ}$.

A second lead collimator (1.7$^{\circ}$ aperture) was placed in front of a 53.3 mm diameter, 40.5 mm long intrinsic germanium (HPGe) detector, used to capture the Compton-scattered gamma. Careful alignment of both collimators was performed at $\Theta_{C}=0^{\circ}$. Fig.\ 3 contrasts the energy loss to electrons in NaI[Tl] with its Klein-Nishina expectation, as a function of $\Theta_{C}$.

\begin{figure}[!htbp]
\includegraphics[width=0.46\textwidth]{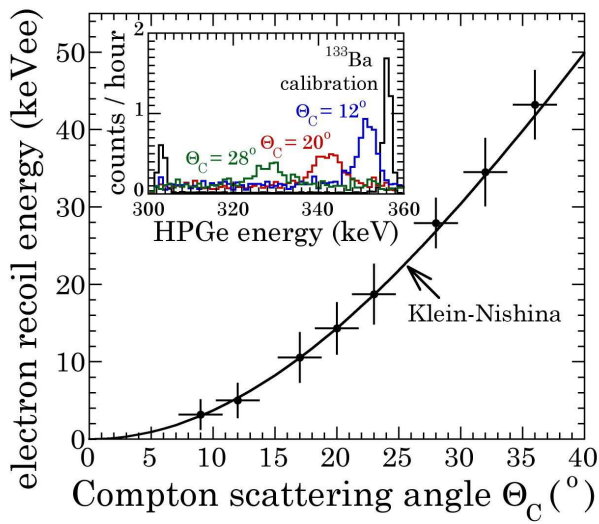}
\caption{\label{fig:CoGeNT-Traces} Difference between $^{133}$Ba incident gamma energy (356 keV) and the centroid of the distribution of HPGe signals coincident with NaI[Tl] events (i.e., energy deposited in NaI[Tl]  via Compton scattering), plotted against $\Theta_{C}$. Error bars are not statistical: horizontal bars correspond to the aperture of the second lead collimator, and vertical to the standard deviation in the HPGe signals. Their overlap with the Klein-Nishina expectation is witness to the good angular alignment obtained in this setup. {\it Inset:} Example spectra of HPGe signals coincident with NaI[Tl]. Two peaks used for HPGe energy calibrations are shown. These were performed prior to each $\Theta_{C}$ measurement, to monitor HPGe gain stability.}
\end{figure}

Binary data from the Acqiris digitizer were read and analyzed using a LabVIEW code.  Fig.\ 4 shows the sum of all Compton measurements. A band of coincident HPGe-NaI[Tl] signals is clearly discernible against residual backgrounds, following the application of cuts to the data (removal of events with significant NaI[Tl] scintillation present already prior to coincidence time, and of NaI[Tl] signals with mean scintillation decay times outside of the 50-500 ns range). The current at the NaI[Tl] PMT was integrated over the 3 $\mu$s following an interaction, and compared to the single photoelectron (PE) mean current (Fig.\ 8 inset), in order to express the NaI[Tl] energy deposition in number of PE. The electron recoil energy deposited in NaI[Tl] was obtained from the difference between the incident 356 keV gamma energy and the energy registered in the HPGe (horizontal axis in Fig.\ 4). The inset in Fig.\ 4 displays the normalization of these results to PE yield per keVee imparted to an electron recoil. The expected nonlinearity \cite{nonprop1, nonprop2} is observed. The quantum efficiency of the UBA PMT is estimated at $\sim$33\% for the NaI[Tl] spectrum of light emission: accounting for the $\sim$25\% reduction in light yield expected for 1 MeV gammas \cite{nonprop1}, this result is in rough agreement with expectations (47 scintillation photons/keVee here vs.\ 40 photons/keVee typically quoted for NaI[Tl] at 1 MeV). Given the good agreement of the observed PE yield with the corresponding curve in \cite{nonprop1} (Fig.\ 4, inset), this last is adopted to extrapolate present measurements to the interval 50 - 200 keVee, via the expression PE/keVee = $16.65 + 3.81\times e^{-0.0143 \times keVee}$.

\begin{figure}[!htbp]
\includegraphics[width=0.46\textwidth]{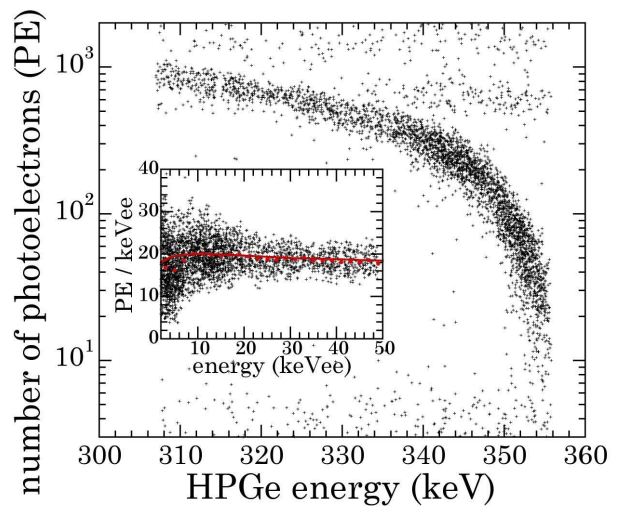}
\caption{\label{fig:CoGeNT-Traces} Number of PE generated by Compton-scattered electrons in NaI[Tl] vs.\ energy registered in HPGe (see text). {\it Inset}: NaI[Tl] photoelectrons per keVee measured in the 2-50 keVee range. Red dots correspond to the mean value computed for 2 keVee bins. The red line is the non-linear behavior described in \protect\cite{nonprop1}, normalized to these data at 10 keVee.}
\end{figure}

\section{Neutron Scattering Measurement}

The arrangement in Fig.\ 2 is converted to that in Fig.\ 5 by simple removal of HPGe detector, gamma source and collimators, and switching of two cables. The same goniometric stage used to hold the gamma source is utilized to mount a Bicron 501A liquid scintillator cell (5 cm diameter, 5 cm long), able to distinguish neutron from gamma interactions via pulse-shape discrimination (PSD) \cite{psd} (Fig.\ 6). This cell is used to detect neutrons scattered off NaI[Tl] at an angle $\Phi_{n}$. It is also optimal in its decay time characteristics (3.2 ns) for fast-coincidence applications. Its PMT was surrounded by mu-metal:  tests were performed to ensure that no measurable change in PMT gain occurred with $\Phi_{n}$. A large tank of borated water was used to block D-D neutrons from reaching the cell following a direct path. The distance between 501A cell and NaI[Tl] was varied within the interval 20 - 45 cm, using the shortest distances for large values of $\Phi_{n}$, in order to compensate for the smaller number of scattered neutrons expected there, at the expense of some loss in angular resolution.

\begin{widetext}

\begin{figure}[!htbp]
\includegraphics[width=1.\textwidth]{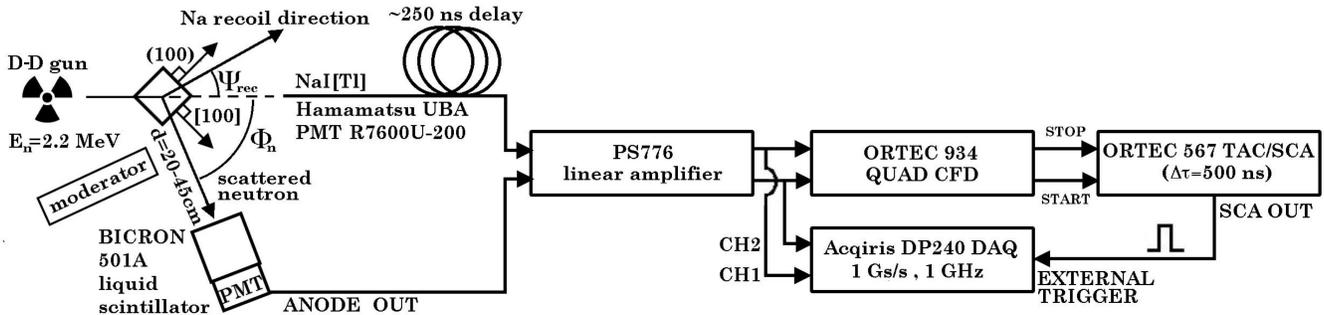}
\caption{\label{fig:electronics} Experimental setup used for neutron scattering measurements (see text). The long (vertical) axis of the NaI[Tl] crystal points into the plane of the figure. A thin lead foil, not shown, is used to block low-energy x-rays emanated by the D-D generator. The distance between deuterated target plane and NaI[Tl] was 42 cm. The PS776 is a 16 channel amplifier.}
\end{figure}

\end{widetext}

\begin{figure}[!htbp]
\includegraphics[width=0.46\textwidth]{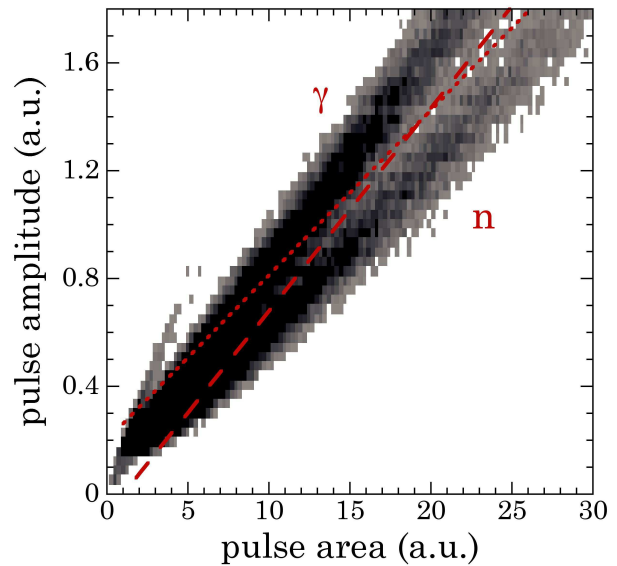}
\caption{\label{fig:CoGeNT-Traces} PSD against gamma backgrounds affecting the Bicron 501A cell \protect\cite{dan2,psd}. Events below the dotted line are passed by the off-line analysis, maximizing the acceptance of scattered neutrons at the expense of some gamma contamination. More stringent cuts (dashed line) are used for measurements of CsI[Na] quenching in this same setup \protect\cite{cosi}.}
\end{figure}

 An important detail noticeable in Fig.\ 5 is the trigger configuration. The coincidence sequence between 501A and NaI[Tl] detectors is armed by events in the first, and closed by events in the second, i.e., in time-reversed order.  To compensate for this, the NaI[Tl] signal is delayed by $\sim$250 ns, and the TAC/SCA used to determine coincidences is set to a sufficiently long window (500 ns). Digitizer trace length and external trigger position within are set to capture the entire sequence of events. The reason for this counterintuitive arrangement is the long decay time of NaI[Tl] scintillation ($\sim$200 ns), when compared to the time-of-flight (TOF) for $\sim$2.2 MeV scattered neutrons traveling between the two detectors ($\sim$15 ns, Fig.\ 7). A standard trigger arrangement where NaI[Tl]  events arm the TAC/SCA results in many missed triggers for low-energy (few PE) NaI[Tl] signals where no PE is generated within the first 15 ns following a nuclear recoil. This systematic effect is entirely bypassed by the present arrangement.

\begin{figure}[!htbp]
\includegraphics[width=0.5\textwidth]{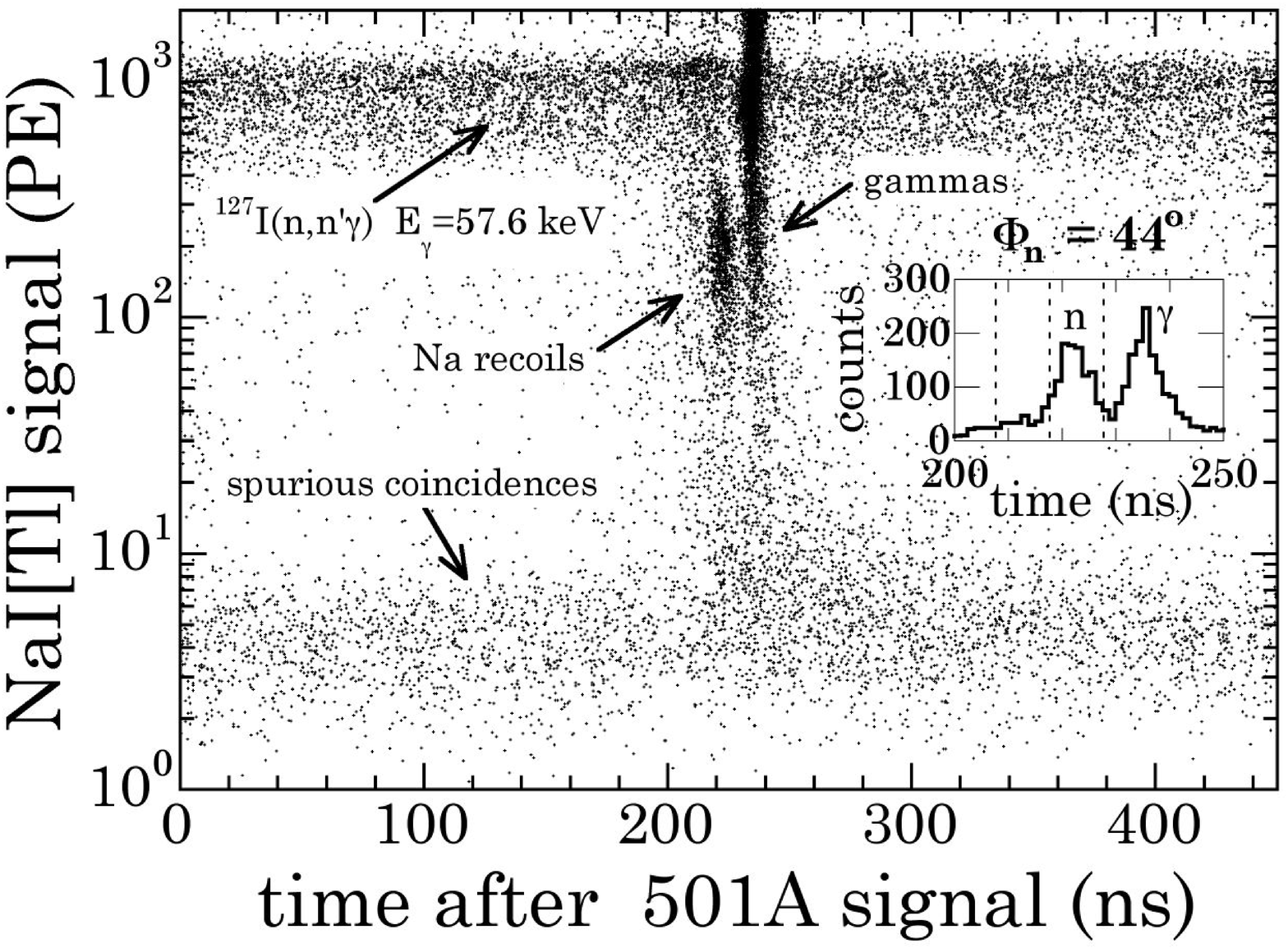}
\caption{\label{fig:CoGeNT-Traces} Events passing all cuts in neutron scattering runs taken at $\Phi_{n}=44^{\circ}$, depositing an average 54 keVnr in NaI[Tl]. {\it Inset}: distribution of events in the $10 <$ PE $< 400$ interval, projected onto the horizontal time axis. The expected TOF of 15 ns for a $E_{n}=2.25$ MeV neutron traversing a distance d = 30 cm between NaI[Tl] and 501A detectors is observed \protect\cite{chag}. See text for a discussion on trigger configuration.}
\end{figure}

NaI[Tl] PMT current integration is performed identically for Compton scattering data, single PE spectrum reconstruction (Fig.\ 8, inset), and neutron scattering runs. This is important, given that the quenching factor is determined by the ratio of these currents (expressed in number of PE equivalent) for comparable electron and neutron recoil energies. To this end, any small fluctuations in the DC-level of NaI[Tl] PMT traces are accounted for in the analysis, PMT current integration times are kept constant for NR and ER, etc. The experimental arrangement emphasizes, by design, keeping changes to a minimum in going from electron scattering to neutron scattering measurements.  Background cuts during the analysis of neutron recoil data are the same as in section III, with the addition of those described in Fig.\ 6. The resulting events passing cuts display a clear separation between nuclear recoils and residual NaI[Tl] - 501A coincidences mediated by numerous thermal capture background gammas (Fig.\ 7).

 \section{Extraction of the Quenching Factor} 
 
Three vertical dotted lines in the inset of Fig.\ 7 define two time domains of equivalent span, the one on the left containing a background of spurious events, the one on the right dominated by nuclear recoil signals in NaI[Tl]. A spectrum of NaI[Tl] light yield (number of PE) can be formed for events in each of these time domains, for each scattering angle probed. Subtraction of the background spectrum from the recoil-dominated spectrum generates the residuals in Fig.\ 8. The centroids of Gaussian fits to these provide the mean number of PE generated by nuclear recoils in NaI[Tl] at each $\Phi_{n}$. The corresponding mean nuclear recoil energies (horizontal axis values in Fig.\ 9) are extracted from MCNP-PoliMi \protect\cite{polimi} simulations of the runs. These energies are in good agreement with expectations from basic neutron kinematics \protect\cite{minowa}. The ratio between these PE centroids and the PE yield observed for electron recoils of the same energy (Fig.\ 4, and extrapolation beyond 50 keVee discussed in section III) is identified as the sought quenching factor. This is shown in Fig.\ 9.

 \begin{figure}[!htbp]
\includegraphics[width=0.46\textwidth]{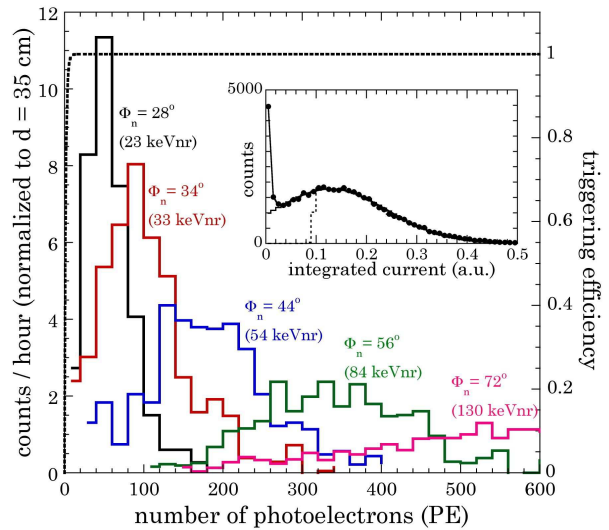}
\caption{\label{fig:CoGeNT-Traces} Example distributions of sodium recoil signals in NaI[Tl], labelled by $\Phi_{n}$ and mean recoil energy. A dotted line represents the calculated triggering efficiency (see text). {\it Inset}: distribution of single photoelectron (SPE) current in the NaI[Tl] PMT. A dotted histogram represents the fraction of these SPEs (69 \%) triggering the constant fraction discriminator (CFD, Fig.\ 5). The stability of this fraction was monitored throughout the experiments.}
\end{figure}

\begin{figure}[!htbp]
\includegraphics[width=0.46\textwidth]{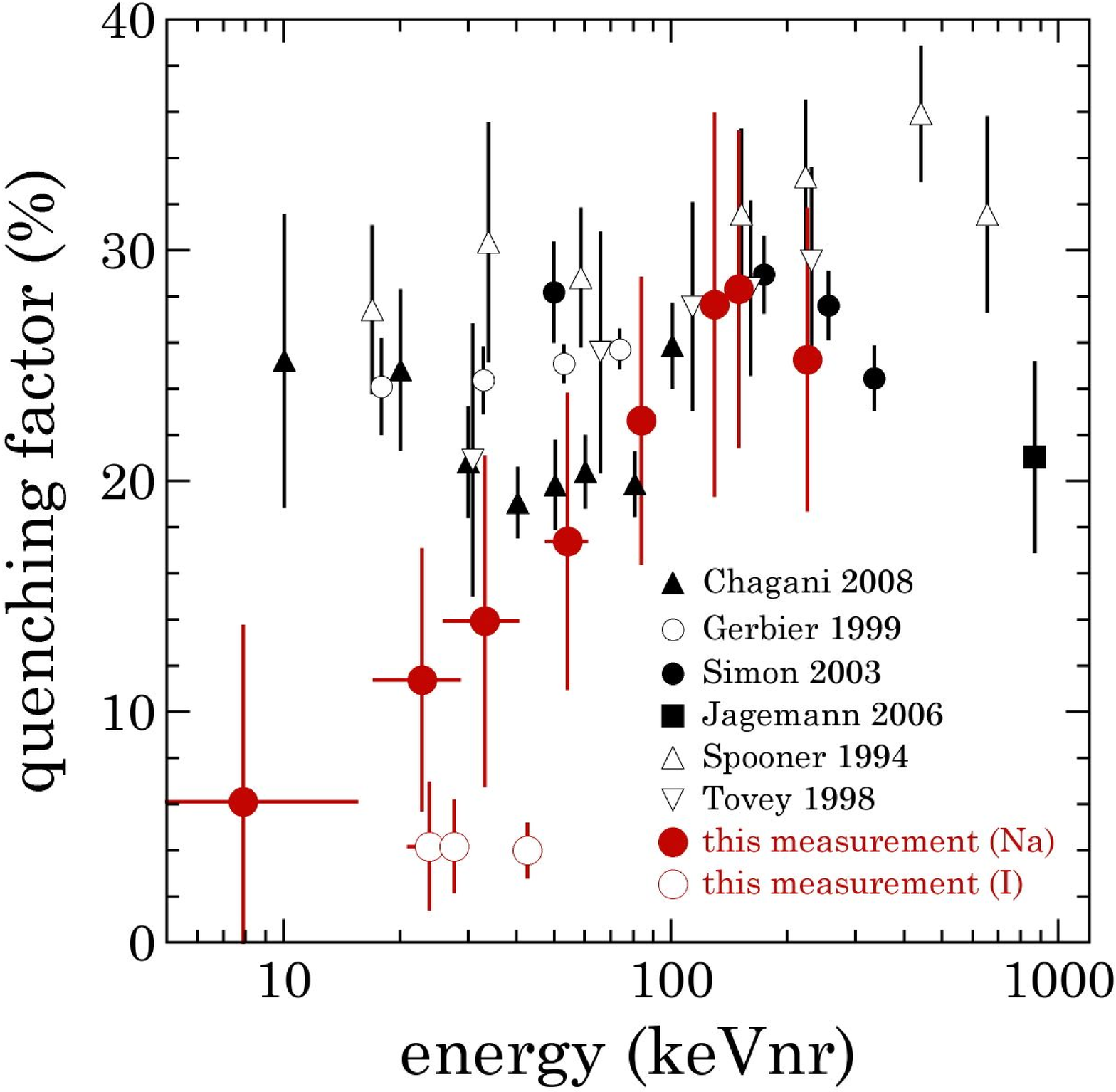}
\caption{\label{fig:CoGeNT-Traces} Quenching factor for Na and I recoils in NaI[Tl], compared to previously obtained values \protect\cite{chag,jagemann,simon,gerbier,tovey,spooner}. Horizontal error bars correspond to the  dispersion in simulated $E_{r}$, vertical to the dispersion in the experimental distributions in Fig.\ 8.}
\end{figure}

Fig.\ 8 also displays the calculated triggering efficiency as a function of number of PE generated by a neutron recoil in NaI[Tl]. Formally, this is computed as the probability of a binomial distribution with PE trials and compound success probability $p = p_{1} p_{2}$ returning a number of successes equal or larger than one, where  $p_{1}=0.69$  is the probability of an individual PE triggering the CFD (Fig.\ 8, inset) and  $p_{2}=1-exp(-\Delta t/\tau)$. Here $\Delta t = 230$ ns is the time window available for a PE to set off the trigger configuration described above (see Fig.\ 7) and $\tau \sim$ 200 ns is the decay time for NR-induced scintillation (Fig.\ 10). In other words, $p_{2}$ accounts for the finite probability that the first PE generated in the NaI[Tl] is not prompt enough to stop the TAC/SCA within its 500 ns range, after accounting for the effect of the delay loop. The effect of $p_{2}$ on reducing the triggering efficiency for a given number of PE is more pronounced for slower (longer $\tau$) scintillators such as CsI[Na] \cite{cosi}. 

As computed, this triggering efficiency reaches unity for PE $\gtrsim$ 10, i.e., only the PE spectrum corresponding to runs at the smallest scattering angle $\Phi_{n}\!=16^{\circ}$ ($E_{r}\!\sim\!8$ keVnr) requires a correction based on it. That the rest of the measurements collected at larger scattering angles are free from any such threshold effects can be ascertained by observing the left shoulders of the distributions in Fig.\ 8, which are not similarly shaped by a triggering  efficiency curve (see discussion around Fig.\ 4 in \cite{xecritique1}). 

Iodine recoils emerged distinctly above the few PE spurious coincidence noise (Fig.\ 7) only at the three largest values of $\Phi_{n}$ attempted ($72^{\circ}$, $78^{\circ}$, $102^{\circ}$). Given the proximity of their PE distributions to the triggering efficiency threshold, these were all corrected by the efficiency curve prior to extracting a Gaussian best-fit. The iodine quenching factor obtained is considerably smaller than for sodium recoils, similarly to previous results \cite{damaquenching,fushimi}, and comparable to values obtained for Cs and I recoils in CsI[Na] in this same setup \cite{cosi}.

\begin{figure}[!htbp]
\includegraphics[width=0.46\textwidth]{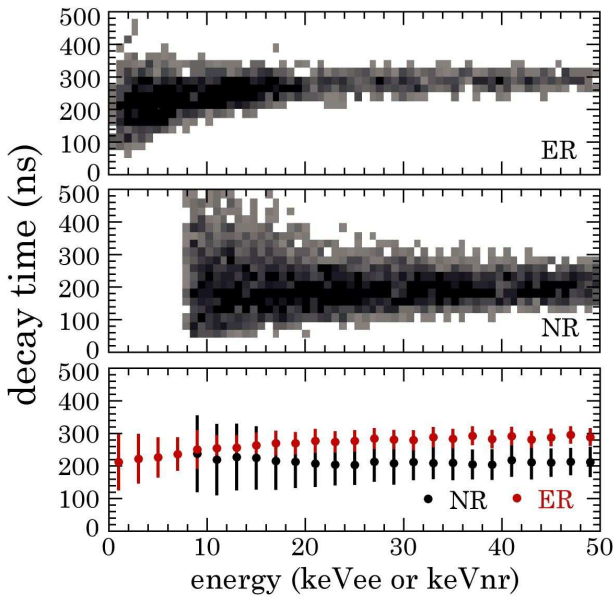}
\caption{\label{fig:CoGeNT-Traces} Grayscale intensity plots showing the observed mean decay time of NaI[Tl] scintillation. NR energies are in keVnr, ER in keVee. A single exponential decay component is used in fits to the running integral of PMT currents. The bottom panel shows the centroids and dispersion of the data in 2 keV bins. The evolution of these centroids with energy is very similar to that found in previous work \protect\cite{chag,gerbier,tovey,smithdecaytime}.}
\end{figure}
 
\section{Search for an Ion Channeling Effect}

The goniometric stage holding the NaI[Tl] detector can be used to select an orientation allowing a recoiling sodium ion to (in principle) channel down the (100) plane perpendicular to the known [100] growth axis of the crystal (Fig.\ 5 and inset of Fig.\ 11). The corresponding angle $\Psi_{rec}$ varies with $\Phi_{n}$, and can be calculated using basic kinematic relations \cite{minowa}. In \cite{prochannel} it was postulated that such a channeled ion should exhibit a $Q_{Na}\sim$1, i.e., lose energy exclusively through ionization much like an electron, and not via secondary nuclear recoils. In \cite{conchannel} it is claimed that "blocking" effects should render this impossible for recoiling ions originating in nuclei initially at rest on the lattice, i.e., the case for WIMP or neutron interactions. 
 
\begin{figure}[!htbp]
\includegraphics[width=0.46\textwidth]{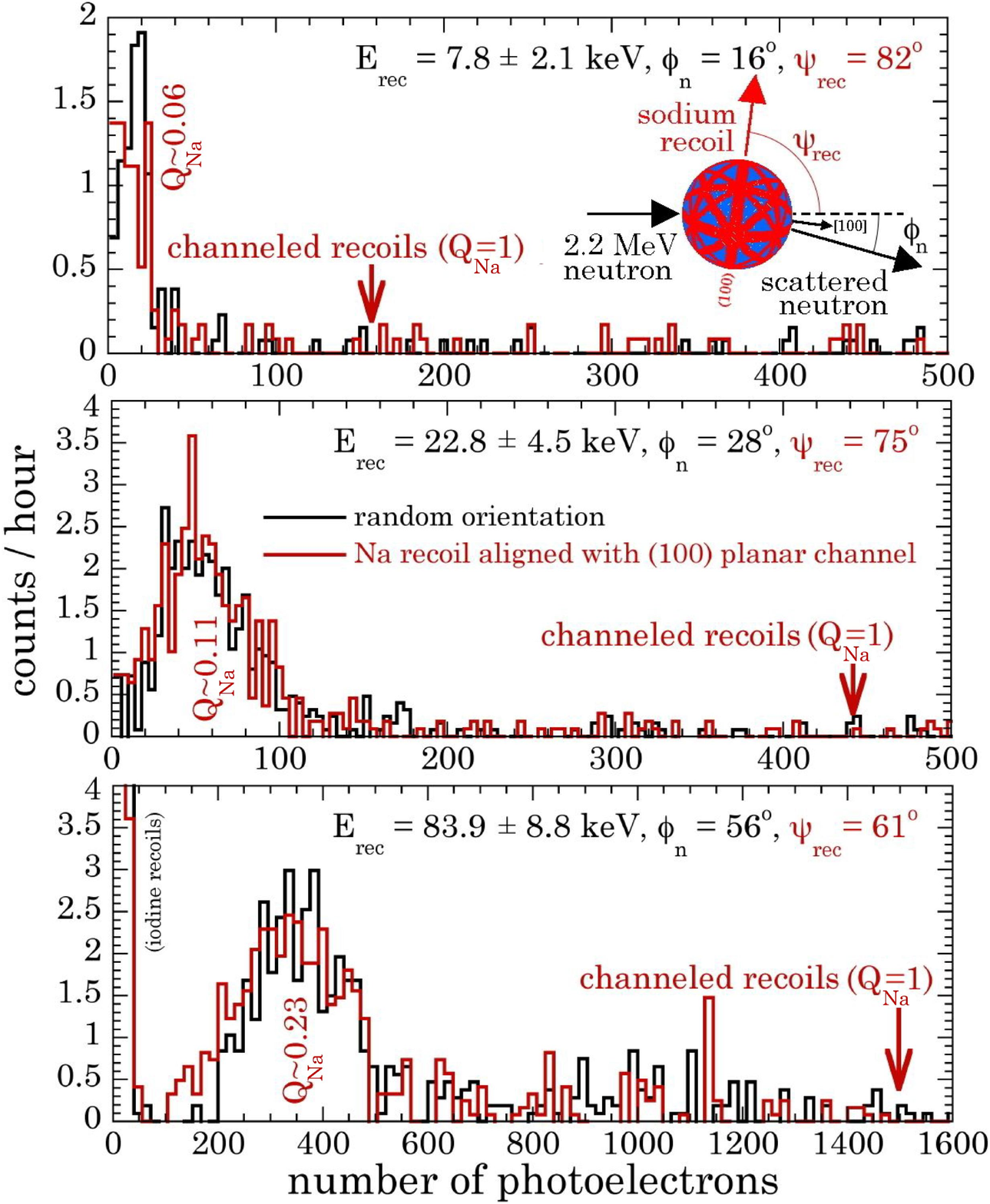}
\caption{\label{fig:CoGeNT-Traces} Results from a negative search for sodium recoil channeling in single-crystal NaI[Tl] (see text).  "Random orientation" corresponds to an alignment of the [100] crystal axis with the direction of the incoming neutrons, i.e., an orientation for which no significant channeling would be expected. The inset displays a HEALPix \protect\cite{conchannel,nassim,healpix} map of available channels in the NaI[Tl] lattice, represent by red bands. These are to scale for $E_{r}=10$ keVnr (see text).}
\end{figure}

The angular precision of the present setup cannot be claimed to be better than $\sim1^{\circ}$. Similarly, the manufacturer of the NaI[Tl] scintillator listed the tolerance of the requested crystal orientation to be  "a few degrees".  Nevertheless, the angular dimension of the (100) planar channel would be sizable for low-energy sodium recoils. For example, in NaI[Tl] this room-temperature channeling angle should be $\sim\!\pm5.6^{\circ}$ around the (100) plane for  $E_{r}\!=\!10$ keVnr, $\sim\!\pm4.0^{\circ}$ at $E_{r}\!=\!28$ keVnr, and $\sim\!\pm3.5^{\circ}$ at $E_{r}\!=\!41$ keVnr \cite{nassim}. Taking into consideration the effect of the finite solid angle subtended by the 501A cell as seen from the position of the NaI[Tl] scintillator, the measurement attempted here should be forgiving enough to display some evidence for ion channeling, if the process was possible at few keVnr for nuclei knocked off their positions in the lattice.

Fig.\ 11 illustrates the negative results from this search. Vertical arrows mark the expected position of the light yield distribution for sodium recoils if channeling was realized, i.e., for $Q_{Na}\!=\!1$. No difference was observed between control runs ("random orientation") and runs with values of $\Psi_{rec}$ selected to favor channeling along (100). For $\Phi_{n}\!=\!28^{\circ}$ additional runs at  $\Psi_{rec}\!=\!72^{\circ}$ and $\Psi_{rec}\!=\!78^{\circ}$, i.e., a few degrees beyond the nominally  favored $\Psi_{rec}\!=\!75^{\circ}$, were performed. This was done to account for the possibility of some significant misalignment in the setup. These runs returned spectra similar to those shown in the figure. 

Following these measurements, an attempt was made to test the crystallographic alignment of the crystal using x-ray scattering at Argonne National Laboratory. The presence of Laue spots confirmed the single-crystal nature of the sample, but it was impossible to obtain information beyond the few degree tolerance listed  by the manufacturer (x-ray penetration through the scintillator casing was minimal, and its removal led to the expected rapid degradation of hygroscopic NaI[Tl] surfaces).  However, for the reasons listed above, and barring a neglect by the crystal manufacturer to adhere to alignment instructions, the absence of any excess at $Q_{Na}\!=\!1$ in the red histograms of Fig.\ 11 suggests that the arguments against channeling in \cite{conchannel} are now experimentally confirmed.

\section{Implications for Dark Matter Searches} 

While the decreasing $Q_{Na}$ obtained in this experiment challenges several previous results, the author has confirmed a value of $Q_{Na}\!\lesssim\!0.1$ for recoils below 24 keVnr using a new independent technique involving an $^{88}$Y/Be  photo-neutron source. This is treated in a separate publication \cite{ybe}. As discussed in section II and \cite{xecritique1}, the combination of poor light yield and lack of sufficient control of systematics near threshold can produce a false impression of constant or increasing quenching factors with decreasing recoil energy. Based on present experimentation, this seems to have been the case for previous measurements of $Q_{Na}$ below $E_{r}\sim40$ keVnr, from which a discussion of threshold effects is notoriously absent (specifically, control of expected vs.\ observed event rates at recoil energies strongly impacted by triggering efficiency  \cite{xecritique1}). In recent times, improvements to the methodology of quenching factor measurements for LXe \cite{manamana,plante} have clarified a similar situation, one that led to an interpretation of dark matter search data \cite{1stxenon} based on markedly optimistic quenching factors \cite{danjuan}. New quenching measurements performed by the author on CsI[Na] \cite{cosi} indicate that this unfortunate situation may also extend to previous studies of this other scintillator \cite{kims}. 

\begin{figure}[!htbp]
\includegraphics[width=0.46\textwidth]{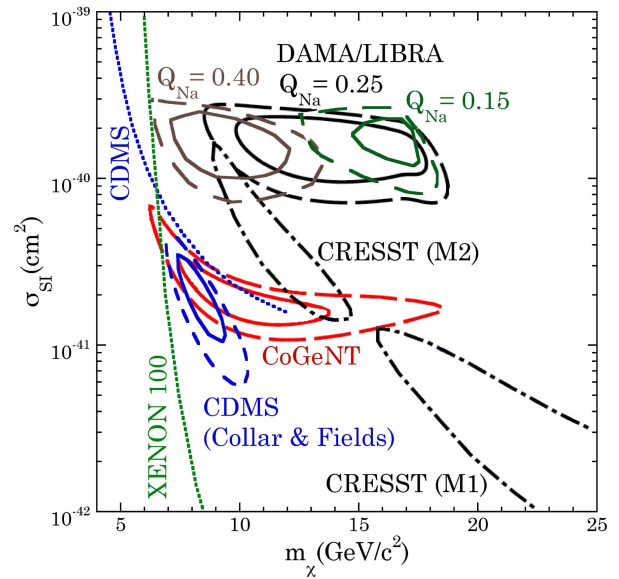}
\caption{\label{fig:CoGeNT-Traces} Effect of $Q_{Na}$ on a DAMA/LIBRA ROI, adapted from \protect\cite{kelso}. XENON100 exclusions have present limited credibility for $m_{\chi}\lesssim12$ GeV/c$^{2}$ \protect\cite{gerbier2}. A CDMS ROI is from \protect\cite{mlcdms}.}
\end{figure}

Fig.\ 12 encapsulates the impact of present measurements specifically on DAMA/LIBRA. A value of $Q_{Na}\sim0.1$ for putative WIMP recoils leading to energy depositions near the 2 keVee DAMA/LIBRA threshold displaces the region of interest (ROI) for a WIMP interpretation well into the realm excluded by several other techniques, significantly increasing the existing tension.  While deviations from the assumption of a Maxwellian velocity distribution for the local galactic halo can still reconcile this displaced DAMA/LIBRA ROI with CRESST and CoGeNT anomalies \cite{kelso,neal}, this is made more difficult by present results. It can be argued that these halo velocity deviations are to be expected \cite{halo}: the upcoming availability of an improved catalog of nearby stellar velocities from the GAIA satellite probe should be able to remove these uncertainties \cite{carlos}. 

The effect of ion channeling would have been to displace the DAMA/LIBRA ROI towards smaller values of WIMP mass and lower couplings \cite{prochannel}, i.e., precisely towards agreement with other anomalies, and away from constraints from other techniques. The objections raised in \cite{conchannel} and the present absence of any indication for this process leave little room for this possibility (notice however the possible caveat listed above). 

In view of the state of affairs, our emphasis should be on continuing to improve the quality of these critical low-energy calibrations \cite{ybe}. Efforts \cite{sorensen} to interpret the microscopic physics underlaying existing (but questionable) calibrations may be premature. If the trend towards vanishing quenching factors at few keVnr continues, scintillating materials may prove to be of limited to no use in searches for low-mass ($\lesssim10$ GeV/c$^{2}$) WIMPs.

\begin{acknowledgments}
The author is indebted to N. Fields, D. Hooper, T. Hossbach and N. Weiner for many useful conversations, N. Bozorgnia, G. Gelmini and P. Gondolo for encouragement and much important information about channeling, and F. Bartolome, M.L. Sarsa and X. Huang for suggesting and facilitating x-ray scattering measurements at ANL.  The author acknowledges the Aspen Center for Physics, supported by NSF Grant 1066293, for hospitality during the completion of this manuscript. 
\end{acknowledgments}

\end{document}